# The 'Earth Rocket': A Method for Keeping the Earth within the Habitable Zone

Mark A. Wessels, Ph.D.
Collin County Community College, Preston Ridge Campus
November 22, 2017

**Abstract**
The Sun is expected to increase its radiant output by about 10% per billion years. The rate at which the radius of Earth's orbit would need to increase in order to keep the present value of the Sun's radiant flux at the Earth *constant* is calculated. The mechanical *power* required to achieve this is also calculated. Remarkably, this is a small fraction (2.3%) of the total solar flux currently intercepted by the Earth. Treating the Earth itself as a rocket, the thrust required to increase the orbit is found, as well as the rate of mass ejection. The Earth has sufficient mass to maintain this rate for several billion years, allowing for the possibility that the Earth could remain habitable to biological life until Sun swallows the inner solar system as a red giant star.

## Introduction

Models of solar physics predict that over time, the Sun will grow hotter over its present state. Its rate of energy production is thought to increase by some 10% every *billion* years[1,2], for at least the next five billion years[3]. At this rate, life on Earth has plenty of time to adjust to new conditions, but within perhaps 2 billion years, the Sun will grow so hot that the oceans themselves will evaporate away into space, leaving the Earth a desiccated, dead world.

Can anything be done to avoid this end, or at least, postpone it? One intriguing possibility could be to expand the size of the Earth's orbit around the Sun. If the distance between the Earth and the Sun could be increased to offset the increase in the Sun's energy output, then conditions for life on Earth could remain stable into the distant future. Earth would remain within the solar system's *habitable zone*, a region around a star where water on a planet can exist in the liquid state – a condition that is, as far as we know, an absolute requirement for life.

Is such a task even possible? Common sense suggests this is an impossible task. The Earth is almost inconceivably massive in comparison with any object that we encounter in our daily lives, and an absolutely astonishing amount of energy would be involved in actually *altering* its present orbit about the Sun. One way to tackle this problem would be to determine how much energy would be required, and then determine if that amount of energy is available at the Earth. The results of this investigation are surprisingly positive.

## Solar Flux Calculations
The total amount of radiated solar power is expected to follow the relation:

$$P(t) = P_0 * (1.1)^t, \qquad \text{(Eq. 1)}$$

where *P(t)* is the Sun's total radiated power as a function of time, $P_0$ is the Sun's current radiated power (3.828E+26 Watts), and *t* is the time from now in *billions* of years. This equation indicates that the Sun's total power output will increase by ten percent with every billion years. This relation is plotted below in Figure 1.

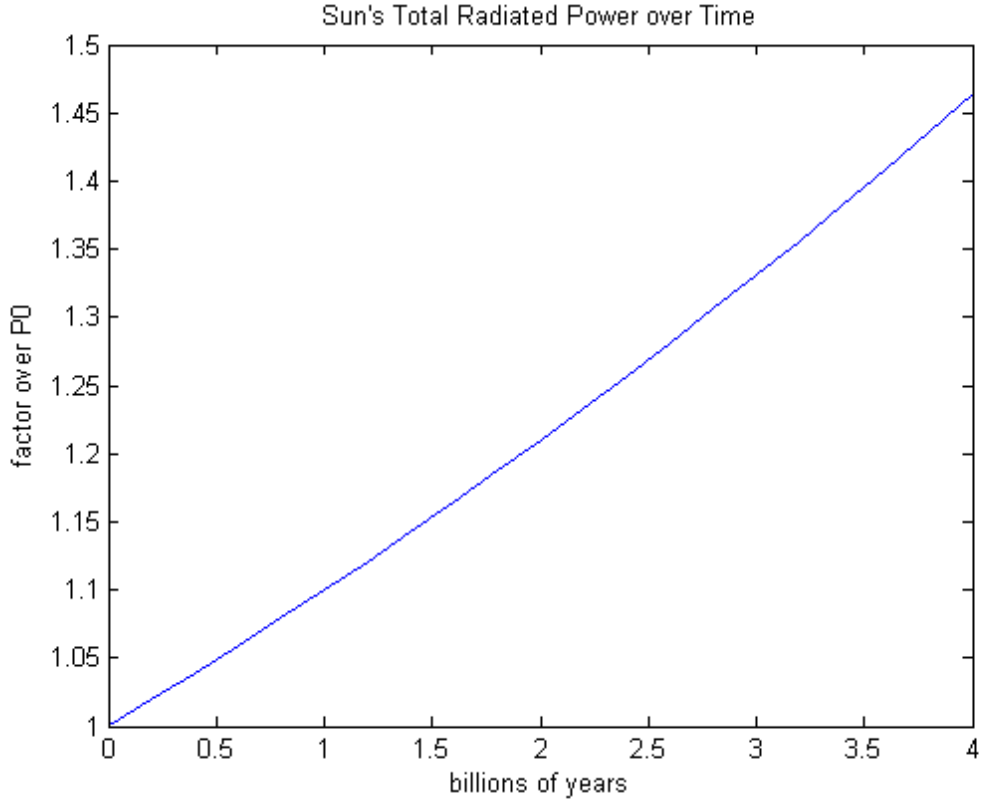

**Figure 1: Sun's Total Radiated Power over Time**. Accepted theories on stellar physics predict the Sun will grow hotter over its remaining life, according to Equation 1. This plot shows the multiplicative increase over the Sun's current power output, some 3.83E+26 Watts.

It is desired to keep the Sun's *radiant flux* (power per unit area) at the Earth *constant* with time. The Sun radiates its power uniformly in all directions. These conditions lead to:

$$P(t) / 4\pi r^2 = P_0 / 4\pi r_0^2 = \text{constant}, \quad \text{(Eq. 2)}$$

where *r* is the orbital radius with time, and $r_0$ is the current radius. Substituting Eq.1 into Eq. 2 and solving for r(t) yields:

$$r(t) = r_0(1.1)^{t/2}. \quad \text{(Eq. 3)}$$

Thus, in order to keep the solar flux received at the Earth constant in time, the Earth's orbital radius r *must increase* as prescribed by Eq. 3. Figure 2 plots this relationship.

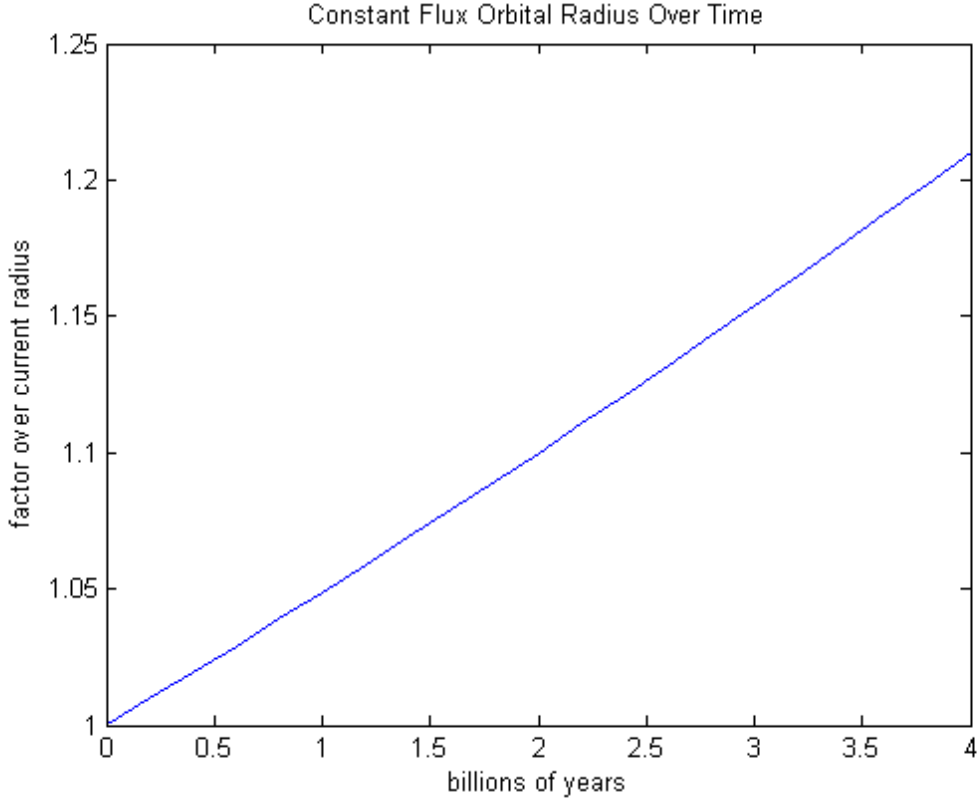

**Figure 2: Constant Flux Orbital Radius Over Time**. In order to keep the solar flux value at the Earth (1360 Watts/m$^2$) constant with time, the size of Earth's orbit must increase according to Equation 3. This plot shows the multiplicative increase over the Earth's current orbital radius.

## Mechanical Power Needed to Increase the Orbital Radius

The amount of power needed to provide for the expansion of the orbital radius can now be calculated. The mechanical energy contained in Earth's (very nearly) circular orbit around the Sun is given by:

$$U(r) = -G*M_{sun}*M_e / 2r, \qquad (Eq.\ 4),$$

where $U(r)$ is the total mechanical energy of the orbit, $G$ is the universal constant of gravitation (6.6732E-11 N.m$^2$/kg$^2$), $M_{sun}$ is the mass of the Sun (1.988E+30 kg), and $M_e$ is the mass of the Earth (5.98E+24 kg). (This quantity is *negative* because the Earth is gravitationally bound to the Sun.) Taking the time derivative of Eq. 4 yields the *power* (time rate of energy increase) needed to increase radius $r$ at a certain rate:

$$P_{mech} = dU(r) / dt = G*M_{sun}*M_e \,((dr/dt) / 2r^2), \qquad (Eq.\ 5)$$

where $P_{mech}$ is the mechanical power and $dr/dt$ is the time rate of change of $r$. Taking the time derivative of Eq. 3 yields the expression for $dr/dt$:

$$dr/dt = \ln(1.1)*(r_0/(2\ t_1))*(1.1)^{t/2}, \qquad (Eq.\ 6)$$

where $ln$ is the natural logarithm function, $t_1$ = 31.56E+15 seconds (the number of seconds in one *billion* years). Substitution of this expression into Eq. 5 and using Eq. 3 to express $(1/r^2)$ yields the expression for the mechanical power (in Watts) required to expand Earth's orbital radius at the constant flux rate:

$$P_{mech} = (G*M_{sun}*M_e/2)*(\ln(1.1) / (2\ t_1 r_0))*(1.1)^{-t/2}. \qquad (Eq.\ 7)$$

In Eq. 7, time $t$ is again measured in billions of years. Eq. 7 is plotted in Figure 3.

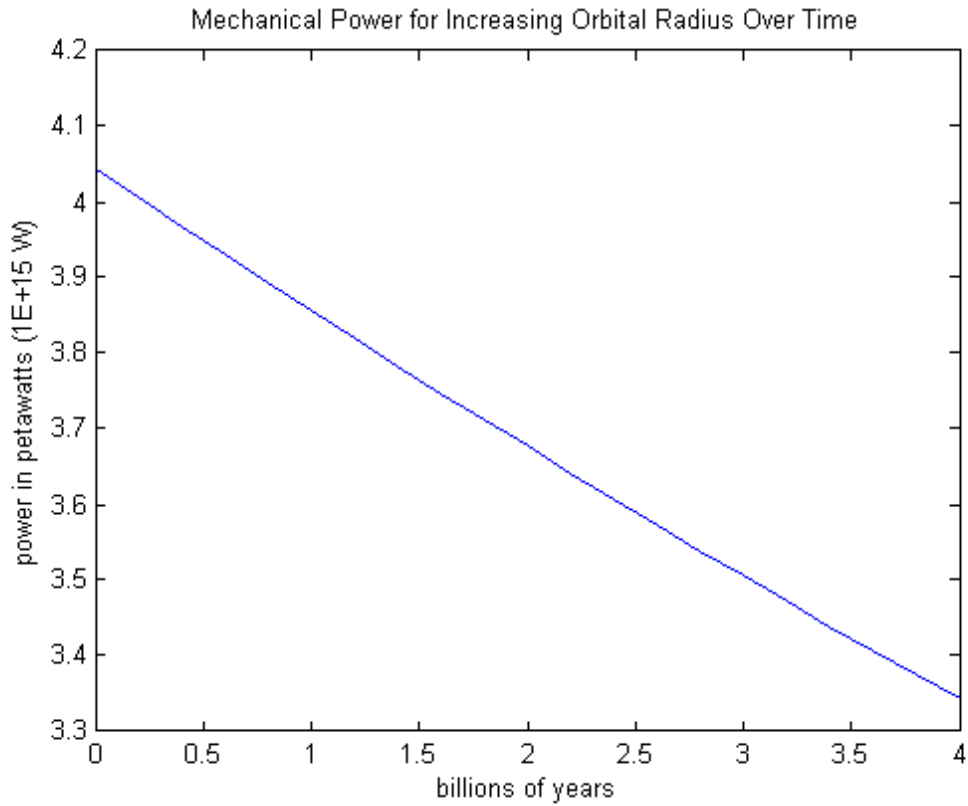

**Figure 3: Mechanical Power for Increasing Radius over Time**. This plot shows how the mechanical power needed to expand the Earth's orbital radius varies with time. Note the power units: *petawatts*. These values are about 100 times greater than all the power currently produced by human civilization. The power *decreases* because the Sun's gravitational strength *decreases* with increasing orbital radius.

## Relationship to Incident Solar Flux

Eq. 7 gives the mechanical power as a function of time needed to keep the incident solar flux at the Earth constant with time. The initial (t=0) value of this power is some 4.04 *petawatts* (4.04E+15 W). The Earth constantly intercepts 1361 W/m$^2$ * $\pi R_E^2$ or *174 petawatts* of solar power ($R_E$ is the radius of the Earth, 6.38 million meters; $\pi R_E^2$ is the *projected* area of the Earth). The 4.04 petawatts is just *2.3 percent of the Earth's total incident solar power*. Even after accounting for inefficiencies present in any practical conversion system, this figure represents only a *minute fraction of the total solar power available at the Earth*, and could therefore provide all the power a system would need for this purpose.

## 'Earth Rocket' Concept

This intercepted energy must now be converted into the work needed to expand the orbit. One possible mechanism for achieving this is a *rocket*. A rocket works by the *Law of Conservation of Momentum* - mass that is expelled in one direction at great velocity produces a reaction force in the opposite. In this case, the rocket would be the Earth itself. Portions of Earth's mass would be ejected behind the Earth at speeds in excess of Earth's escape velocity (11.1 km/sec). (At any lesser speed, these ejected masses would fall back to Earth, negating the rocket action.) The resulting thrust would be coincident with Earth's velocity around the Sun, causing the Earth to spiral slowly outward over time. See Figure 4 below.

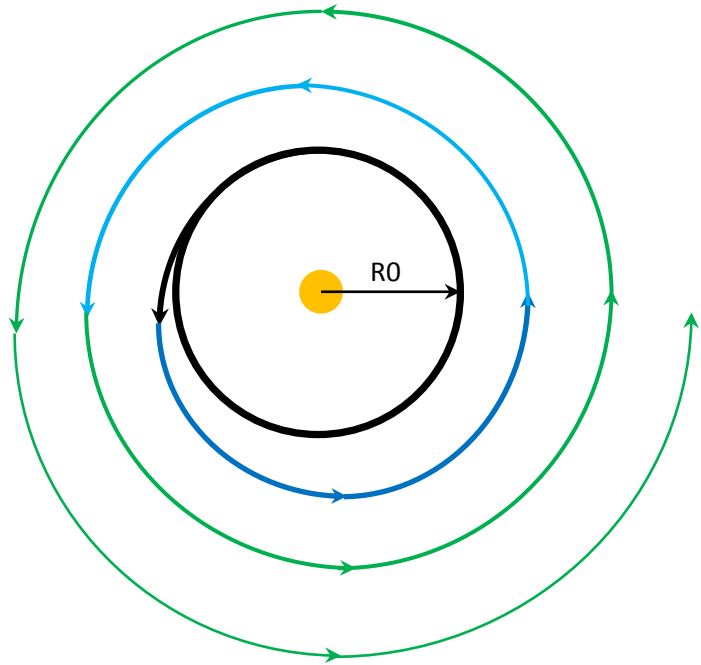

**Figure 4: Outward Spiraling Orbit.** As mass is expelled from the Earth, the resulting thrust adds energy and angular momentum to the Earth's orbit. The result is a slow, outward-spiraling orbit. The arcs shown here in color represent many millions of orbits. The increase in orbital radius would just offset the increase in the Sun's radiant flux over time.

## Thrust Calculation

Consider the work performed by this thrust acting throughout one orbit. In physics, work is defined as *force acting through a distance*. The thrust acts in the direction of Earth's motion around the Sun. Furthermore, its magnitude can be assumed to be nearly constant over any single orbit, because the increase in orbital radius *per orbit* is very small in comparison to the radius of the orbit itself. Thus, the work performed per orbit can be expressed as

$$\text{Work per orbit} \approx (\text{thrust} \times 2\pi r). \quad (Eq.\ 8)$$

But the work per orbit is also equal to the mechanical power x one orbital period (T). Together,

$$\text{thrust} \approx P_{mech} * T / 2\pi r. \quad (Eq.\ 9)$$

Inserting the initial (t=0) values for these quantities yields the value for the thrust at **127.25E+09 Newtons**. This force is equivalent to the liftoff thrust of some 3800 Saturn V rockets[4].

## Mass Ejection Rate

According to the classical rocket equation[5],

$$\text{thrust} = |dm/dt| * v_{exh}, \quad (Eq.\ 10)$$

where *|dm/dt|* is rate of mass ejection, and $v_{exh}$ is the exhaust velocity. Dividing the above value for thrust by the minimum possible exhaust velocity – the escape velocity for Earth – yields the maximum necessary rate of mass ejection: some 11,400 metric *tons per second*. (This figure is approximately that of *four* fully-fueled Saturn V rockets[6].)

## Discussion & Conclusion

These calculations indicate that only a small fraction of the Earth's total intercepted solar power is needed to provide the mechanical work for expanding the size of the Earth's orbit around the Sun, to offset the increasing radiated power that will occur over the Sun's life.

Expelling mass from the Earth itself at great velocities would produce a reaction force – thrust – that would expand the size of the Earth's orbit. In order to produce the necessary thrust, machines capable of hurling over 11,000 metric tons per second, at a speed in excess of Earth's *escape velocity* (11.1 km/sec)[7], would need to be constructed. These ejections would need to continue unrelenting for several billion years into the future. To this author, the only imaginable type of machine even remotely capable of such feats would be a rail-gun of immense size.

Does the Earth have enough mass to sustain this rate of ejection for so long? At this rate, some 380 billion tons would be displaced every year. After one million years, 380 *petatons* (1E+15) will have been ejected. This figure is one part in 15,800 of the Earth's total mass. So, after the first *billion* years (1000 times longer), about one part in sixteen (1/15.8) will have been ejected. (All the previous calculations have held the mass of the Earth and Sun *constant*, but this is a good approximation for at least the first billion years.) After *two* billion years, 1/8 will be gone. After four billion, one-quarter.

These mass fractions seem acceptable. Even with one-quarter of Earth's current mass gone, the Earth would still be very much a planet, still possessing 90% of its current surface gravity (assuming the same average density). This change in surface gravity is insignificant.

Undoubtedly, many such machines, probably hundreds, would need to be constructed all over the planet, working simultaneously. Having so many working at once would reduce the mechanical demands needed of each one. Furthermore, this massive redundancy would help relieve mechanical failures. (If some of the machines were to fail at any given time, the influence would be completely negligible.)

After 4 billion years, the Sun will no longer be the gentle yellow dwarf star it is now. It will have exhausted its hydrogen fusion fuel, and will have mushroomed into a red giant, swallowing the entire inner solar system out to Mars. At that time, the only worlds in this solar system capable of supporting life would likely be the moons of the outer gas giant planets: Jupiter, Saturn, Uranus, and Neptune. Until then, the 'Earth Rocket' could extend life on this planet several billion years beyond its current expiration date.